# Dynamic Tunneling a Prior Phenomenon to Intermediate Resistive States in Defect Engineered Mn-$SrTiO_3$ RRAM Devices

*Gowtham Velpandi‡, Ishaan Tushir‡, Durairaj Murugan†, Thanumalayaperumal Thirunavukarasu and Ranjith Ramadurai**

Department of Materials Science and Metallurgical Engineering, Indian Institute of Technology Hyderabad, Kandi, Sangareddy, Telangana 502285, India.

KEYWORDS: Quantum tunnelling; Resistive switching; Oxygen vacancies; Metal ion defects; neuromorphic computing; RRAM

ABSTRACT

Resistive random-access memory devices work under the principle of resistive switching from high resistive state to low resistive state under applied electric field. In this work, 2%Mn doped $SrTiO_3$, devices were fabricated using PLD in which, the defects are engineered such that the conducting filament formation is expected to have two oppositely charged defects: a) presence of $Mn^{3+}$ dopants, giving rise to a negatively charged $Mn'_{Ti}$ and b) single and double positive oxygen vacancies. This process of two filaments emanating from both the electrodes has shown enhancement in the resistive switching characteristics aiding in forming-free switching that exhibits over $10^2$ as ON/OFF ratio with a memory retention beyond $10^3$ seconds at a low read voltage of 0.1V. All devices demonstrated a stable bipolar resistive switching dominated by conduction filament formation resulting from a synergic effect of oxygen vacancies and the varying oxidation states of Mn based negative defects that propagates toward each other under bias. The formation of defect-based filaments and the lateral segregation of defects around the surface of the bias electrode was evidently observed using x-ray photoelectron spectroscopy of the devices before and after switching. A dynamic tunnelling process is observed as an intermediate process before the conduction filament forms in the SET region and following the snapping in the RESET region. Conduction filaments being heavily doped localizes conduction pathways with a small finite tunnelling barrier gap that are instantaneously formed and ruptures during switching cycles. The dynamic tunnelling behavior gives rise to a tiny spike in current and is observed at the intermediate resistive states which are potential for multi-state usage of these devices. Overall, Mn:STO devices of varying thicknesses show promising retention features and defect engineering offers control over single and/or multi-states memory.

## 1. Introduction

In-memory Computing (IMC) devices has been in high demand for a combination of non-volatile storage and high density specific to AI based applications.[1,2,3,4] Resistive Random-Access Memories (RRAM) has been found suitable for the crossbar architecture and 3D integration facilitating robust IMC pathways.[5,6] RRAM devices are capable of changing resistance upon application of electric pulses which leads to change in the resistive state of the active layer form high resistive state (HRS) to low resistive state (LRS).[7,8] Unlike other memory devices such as magnetic random-access memory (MRAM), ferroelectric random-access memory (FeRAM) that relies on only two states (ON/OFF), RRAM can store data in intermediate resistive states (IRS) besides HRS and LRS, thus greatly improving the data storage density via multi-state resistance to aid the upcoming technological demands.[9,10,11] The most observed mechanism of switching behavior between resistance states are formation of conduction filaments (CFs) and its rupture[11,12,13,14] and interfacial oxidation states variation[15,16,17] with a characteristic abrupt and gradual set/reset respectively. Though RRAM devices show merit towards high density packing, robustness towards radiation and plausible IMC devices, there are technological betterments possible in features like, formation voltage, ON/OFF ratio, long-term endurance, and retention of recorded data etc.[5]

The other aspect of the challenge is to identify a suitable active layer that could be integrated with the existing advanced silicon-based chips. Strontium titanate (STO), a perovskite structure has shown promising feature on being integrated with silicon wafers in both polycrystalline and epitaxial form,[18,19,20] serves as an optimum candidate for RRAM, since it provides flexibility in the type of resistive switching (RS) mechanism based on either Sr-O or Ti-O termination exhibit filamentary or interfacial resistance change,[21] directionality of set and reset polarity influenced by switching voltage[22] as well as pulse width,[23] achieving forming-free voltage via dopants and oxygen vacancy ($V_O^{\cdot\cdot}$) driven defects[12]. Among various dopants such as iron (Fe),[22,24] manganese (Mn),[12,13,25] niobium (Nb),[26,27,28] Mn being an amphoteric dopant,[29] is unique owing to its ability to stabilize in mixed oxidation states ($Mn^{3+}$ and $Mn^{4+}$), predominantly substituting at $Ti^{4+}$ site and facilitates inherent generation of $V_O^{\cdot}$ $and$ $V_O^{\cdot\cdot}$. Thus, assist in formation of p-type ($V_O^{\cdot\cdot}$) and n-type ($Mn_{Ti}'$) forming free CFs that accumulated under electric field.[12]

Investing on the challenges of scalability of RRAM, the significant parameters to influence forming, set, and reset voltages are voltage sweep rate,[23] thickness of the active layer and area of the electrode.[30,31] Forming voltage was reported to be proportional with the thickness of the active layer. Further, thickness dependent $V_O^{\cdot\cdot}$ and its influence on unipolar and bipolar characteristic of RRAM in epitaxial STO has been reported.[15] Moreover, one of the uniqueness of RRAM over its counter memory device is its multistate storage modulated based on voltage window, compliance current which STO also showcases.[24]

Recently, it was observed that the defect engineered Mn doped STO facilitates formation of bi-directional n-type and p-type defects driven CFs and its response to variation in electric field density in polycrystalline 2% Mn doped STO (SMTO) based RRAM fabricated by pulsed laser deposition (PLD). 2% Mn doping concentration was chosen to maximize defect density without compromising the structural integrity of the film.[12,13] This optimal concentration introduces a favorable environment for oxygen vacancies, allowing for precise modulation of vacancy dynamics and filament stability. In addition, the CFs formation is associated with the n-type and p-type defects migrated toward opposite polarity in RS form HRS (filament ruptured) to LRS (filament formed) and vice versa.

Inspite of the established benefits of the RRAM devices the fundamental aspects of understanding the filament formation, models based on number of filaments forming between conductive states etc. has been the focus of research for applicability of these materials for neuromorphic computing and multilevel storage devices.[9,10] In the case of multifilament formation models, the intermediate steps achieved during the process of transition between the LRS and HRS are fascinating and are considered as a potential atomistic phenomenon, that would facilitate multistate accessibility of a given RRAM unit.[32,33] However, if these intermediate resistive states (commonly known as IRS) are a potential feature of the device for neuromorphic computing, the atomistic and charge transport mechanism pertaining to these IRS becomes crucial. In this work, we chose the Mn doped STO thin films which is proven to have a robust filament formation and also possessing features of IRS between the LRS to HRS transition to understand the transport mechanisms that could be potentially involved during the process of formation of IRS. Especially in the case of devices in which CFs formation is associated with the n-type and p-type defects migrated toward opposite polarity during the formation of IRS. In addition, the role of thickness of the devices in formation of these IRS was also studied.

## 2. Experimental Section

### 2.1. Mn-doped STO Growth by PLD

Growth of SMTO thin films was conducted using PLD with a KrF laser that has a wavelength of 248 nm, operates at a repetition rate of 5 Hz, and delivers an energy of 300 mJ per pulse. The target (mentioned in supporting information) to substrate distance is set at 4 cm, with the substrate being a Pt/Si wafer. The process begins with heating the Pt/Si substrate gradually from room temperature to 650 °C at a rate of 10 °C per minute under 10 mTorr. Once the substrate reaches 650°C, the temperature is maintained till the completion of the deposition process. During deposition, laser pulse count was varied to achieve films with different thicknesses ranging from 25 to 870 nm. These pulses enable the stoichiometry transfer of SMTO material onto the Pt/Si substrate, ensuring a uniform and high-quality film. Following the deposition, the temperature is reduced to 100 °C at a rate of 3 °C per minute to prevent any thermal shock to the material. Once this cooling phase is complete, the system is allowed to cool naturally to room temperature[12].

### 2.2. Device Fabrication

After the successful deposition of SMTO on the Pt/Si substrate, the top electrode, made of platinum (Pt) with a thickness of 20 nm and 500 μm diameter, was deposited using the Electron Beam (E-beam) evaporation technique. During deposition, the thickness of the electrode was monitored by a quartz crystal monitor. This method ensures a uniform and precise deposition of the Pt layer, completing the electrode formation. The resulting device exhibits a typical metal-insulator-metal (MIM) sandwich structure with a perpendicular configuration of Pt/SMTO/Pt. This architecture, with platinum as symmetric electrodes and SMTO as the active switching layer (different thickness), forms the basis of the resistive switching mechanism in the RRAM devices.

## 3. Results and Discussion

### 3.1. Structural Characterization

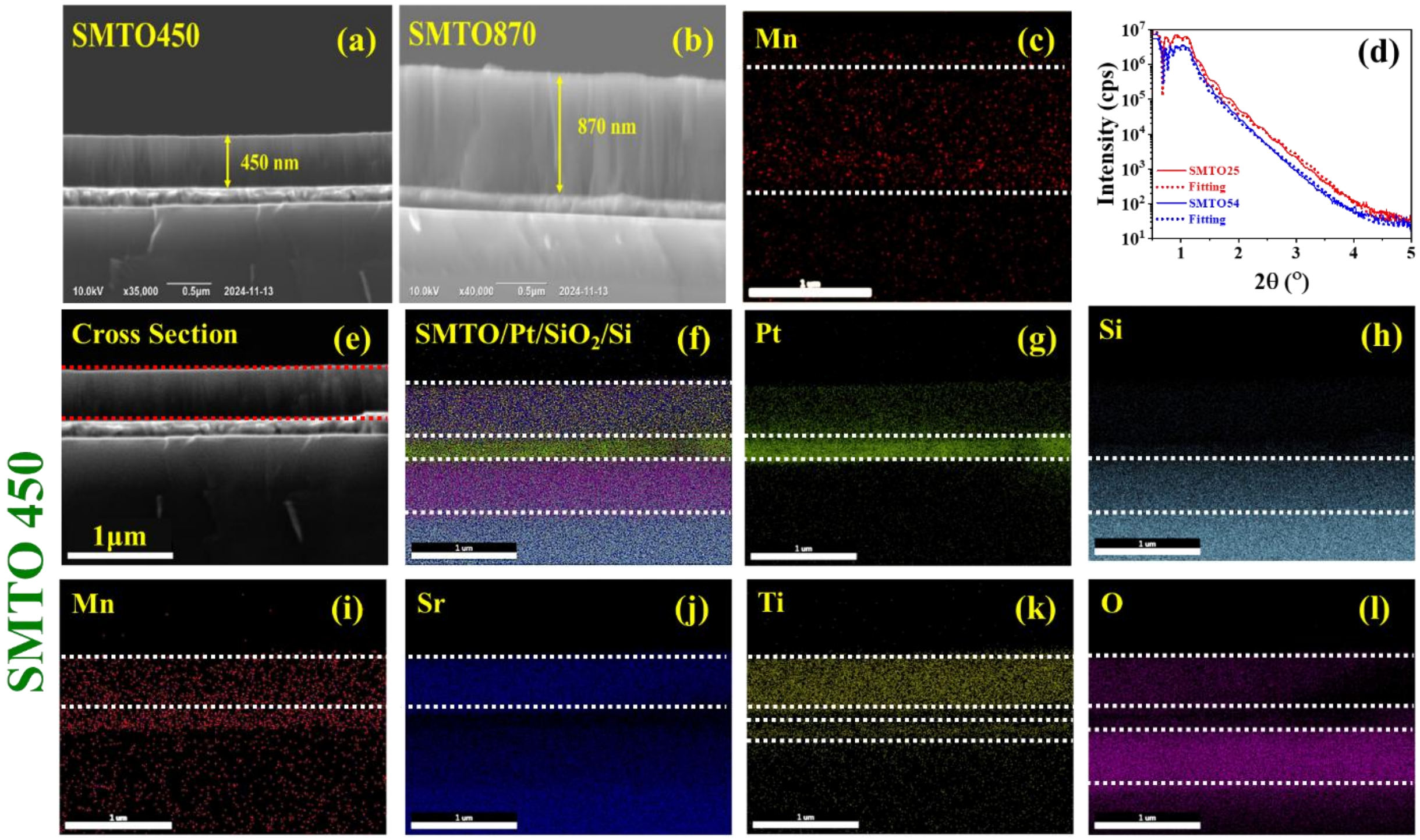

Figure 1. Cross-section FESEM images of (a) SMTO450, (b) SMTO870, (c) EDX mapping of SMTO870 (d) XRR of SMTO25 and SMTO54 and (e-l) EDX mapping of SMTO450

The fabricated RRAM devices by pulsed laser ablation (PLD) were characterized both in terms of structural and electrical aspects. Cross sectional image captured with field emission scanning electron microscope (FESEM) shown in Figure 1(a) and (b) captures the morphology and thickness of the SMTO layers grown on platinized silicon (Pt-Si) substrate with 8000 (SMTO450) and 16000 (SMTO870) pulses respectively. Also, the periodic x-ray reflectivity (XRR) fringes reflected from SMTO layers grown with 500 (SMTO25) and 1000 (SMTO54) pulses extending beyond 4$^{o}$ as in Figure 1(d), confirms the smooth interface quality and surface of the polycrystalline films grown on Pt-Si with homogeneous density and the thickness is calculated to be ~25 and 54nm respectively by XRR fitting of the fringe width. The thickness of SMTO layer scales up linearly with the number of pulses indicating a consistent and precise thickness control over films deposited using PLD with the controlled growth rate of ~0.53Å/pulse estimated from Figure S1. FESEM and XRR outcomes collectively endorses that the deposited SMTO film of thickness ranges from ~25 to 870 nm. Further, the recorded SEM images reveals a dense columnar growth with grains and grain boundaries propagating from the SMTO-Pt interface to the surface of SMTO film. The defects induced via metal ions and $V_O^{\cdot\cdot}$ are anticipated to be localized at the grain boundaries. This uniform columnar morphology achieved is intended to facilitate p-type and/or n-type defects accumulated CFs along the grain boundaries[14,16,34] bridging the top and bottom electrodes providing a predefined pathway eliminating the need of high forming voltage, during the resistance switching response in RRAM. To ensure the homogeneous distribution of Mn dopants, energy dispersive x-ray (EDX) mapping is captured as displayed in Figure 1(e-l). The elemental mapping of Mn, Sr, Ti and O supported with SEM images illustrates a clear distinction between SMTO, Pt, $TiO_2$, $SiO_2$ and Si layers and highlighting the homogeneous distribution of Mn dopants within STO matrix of the PLD grown film in Figure 1 (c and i).

The diffraction patterns in Figure S2(a) confirms the cubic phase of synthesized SMTO target with well-defined distinct crystalline peaks . The peak position of the deposited films of all thickness in Figure S2(b) obtained using Bruker Discover D8 diffractometer in grazing incidence x-ray diffraction (GIXRD) mode, matches the SMTO target and reference in the XRD pattern as shown in Figure S2(a). The relative intensities of the diffraction peaks showed systematic variations with increasing film thickness. These variations imply a thickness-dependent effect on the crystalline nature of the SMTO, likely due to changes in strain relaxation, grain size, or thin film density as the deposition progress. The intense peak observed around 69$^{0}$ in lower thickness samples is due to Si (400) and the intensity had decreased at higher thickness due to reduced penetration through SMTO layer towards the substrate. Sharp and well-defined peaks in the XRD patterns confirm the successful deposition of SMTO thin films with a crystalline perovskite structure. Additionally, the observed thickness-dependent changes highlight the potential to tailor the structural properties of SMTO thin films through precise control of the deposition parameters, offering insights into their optimization for functional applications in electronic and memory devices.

### 3.2. Electrical Characterization

In RRAM, the memory states are switched between HRS to LRS and vice versa by biasing the electric field between set and reset voltage. STO based RRAM are known to show bipolar RS with set voltages at both polarity[22,23] with a defined set voltage during negative bias.[12,13,21] In this study, a probe station (Make: Lake Shore, cryotronics) equipped with two tungsten probes of 25μm tip radius was used to obtain the RS behavior of SMTO active layer of varying thickness with symmetric Pt electrodes, having a MIM device architecture. Direct current (DC) voltage pulse with a pulse width of 2ms was swept in the sequence of 0V → -2V → 0V → +2V → 0V using Keithley 2450 source meter via top electrode while grounding the bottom electrode. The representative data of the observed switching characteristics of SMTO devices of thicknesses ~ 25nm, 110nm and 870nm are given in Figure 2. The data of the rest of the samples studied could be found in supplementary information Figure S3.

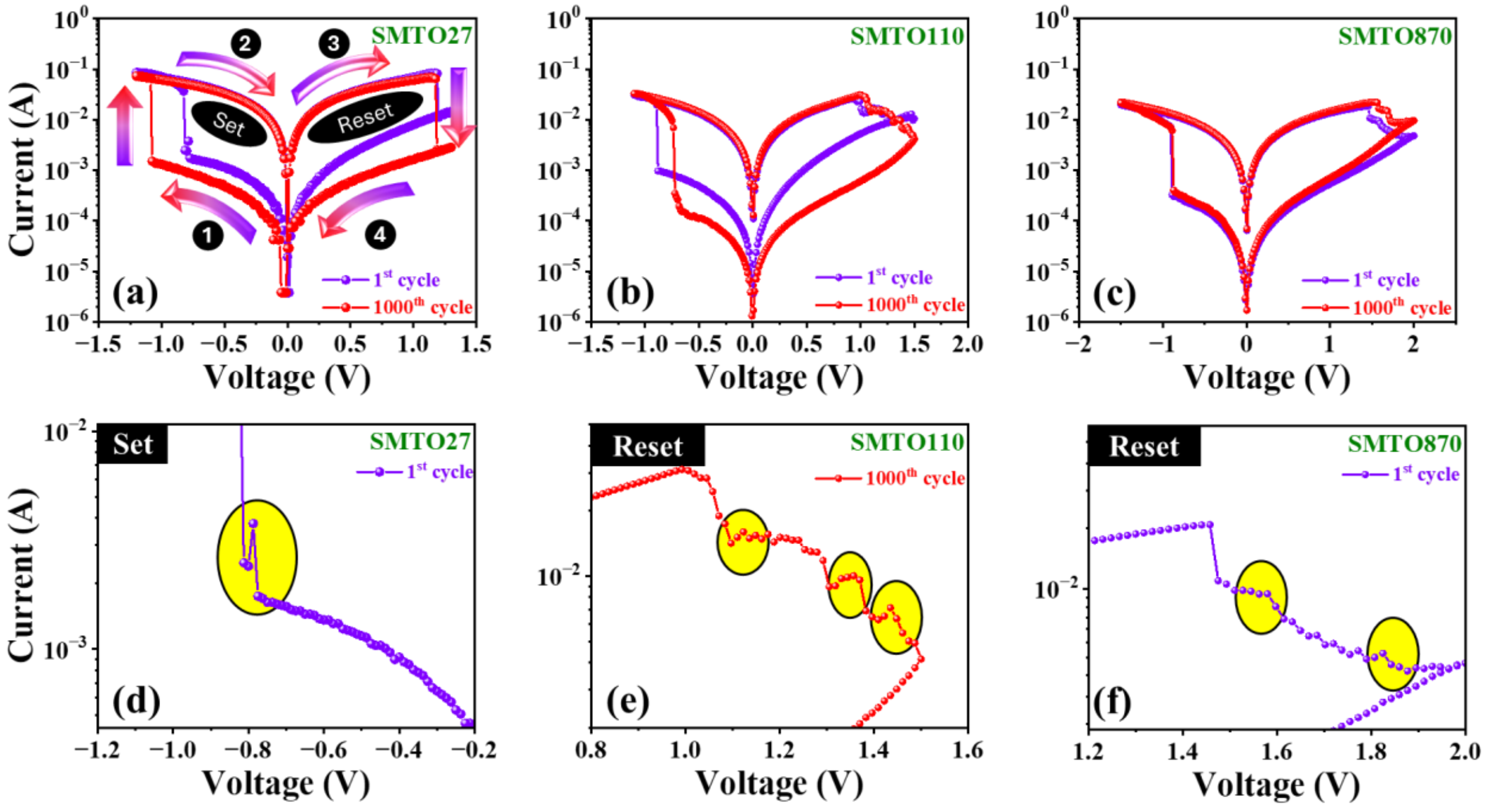

Figure. 2 Semilog I-V characteristics of Pt/SMTO/Pt device with different SMTO thickness of (a) 25nm, (b) 110nm, (c) 870nm and zoomed view of (d) 25nm, (e) 110nm and (f) 870nm highlighting the intermediate tunnelling process during set and reset. (Voltage sweep direction was retained as same for all studies).

In pristine SMTO, the most probable point defects like doubly positive oxygen vacancies ($V_o^{\cdot\cdot}$) and singly negative $Mn^{3+}$ substituted in Titanium site ($Mn'_{Ti}$) are homogeneously distributed. However, under the biased condition, these defects being electrically charged, migrates towards the electrodes with opposite polarity. All fabricated RRAM devices demonstrates bipolar filamentary switching with set (filament formation) in negative bias creating an LRS. When the polarity is reversed, reset occurs in positive bias in which the metal ion and $V_o^{\cdot\cdot}$ disintegrates into the SMTO matrix rupturing the CFs, as observed in Figure 2. This reverts the device back to HRS but not necessarily to the pristine SMTO completely, thus facilitating the subsequent switching between HRS and LRS under cyclic biasing. Such performance enables the utilization of these devices for memory storage and neuromorphic applications of RRAM.

Though STO based RRAM polarity is highly sensitive to the set voltage, all fabricated devices irrespective of thickness exhibited stable bipolar RS with ON/OFF ratio of $10^2$ beyond $10^3$ cycles retaining the resistive hysteresis as seen in Figure 2. Earlier studies on devices with predominant oxygen vacancy driven switching suggests[12] that the positively charge $V_O^{\cdot\cdot}$ migrate towards the electrode that is negatively biased. Thus, in our studies it is possible that $V_O^{\cdot\cdot}$ migrates towards the Pt top electrode during the negative bias and the $Mn_{Ti}'$ defects could accumulate at the bottom electrode. On further bias, the positively charged defect ($V_O^{\cdot\cdot}$ ) and the negatively charged $Mn_{Ti}'$ defects are expected to contribute for the formation of CFs, thereby reducing the barrier layer (SMTO) gap. These CFs thus formed comprising both $V_O^{\cdot\cdot}$ and $Mn_{Ti}'$ creates an equivalence of two dynamic conductive channels meeting withing the device and/or similar to a heavily doped ($p^+/n^+$) channels that connects and transforms the device between HRS and LRS. These CFs merge and recombines to form low resistive Mn (metallic + reduced Mn ion) based CFs, evident from the XPS discussed below which displays the peak corresponding to $Mn^0$ oxidation state. Driven by electric field, these oppositely charged defects ($V_O^{\cdot\cdot}$ and metal ion) based CFs encounter each other, before creating a conduction channel from top to bottom Pt electrode converting the device state from HRS to LRS. Once CFs bridges and recombines, the memory state switches from HRS to LRS and ohmic conduction takes over. Usually, an additional forming voltage higher than the set voltage is applied to facilitate the creation of $V_O^{\cdot\cdot}$ and subsequent CFs formation. But, $Mn^{3+}$ doping at $Ti^{4+}$ site inherently generates $V_O^{\cdot\cdot}$[12]. On the reversal of the biasing known as the RESET phase of the device, the defect migration happens in the opposite directions.

At the instant of formation of these conductive channels, for a given forming voltage a tiny spike in the current could be observed as shown in Figure 2(d). This observation is consistent and observed across over few cycles in the set region of the thinner samples and observed across over 1000 cycles in the reset region of all the devices studied except for the thinnest ones(~25nm). The disappearance of the spike after 10 cycles in samples less than ~ 25nm in the SET region, could be due to the accumulation rate of the defects being relatively quicker than the delay times utilized in this study. Interestingly, these tiny spikes of current were consistently observed in the RESET process of all the relatively thicker films especially at the IRS regions. As mentioned above such a tiny spike in current which is identified as highlighted in in Figure 2 (e and f) are attributed to a short duration dynamic tunnelling (similar in set process) between p-type ($V_O^{\cdot\cdot}$) filament and n-type ($Mn'_{Ti}$ and $Ti'_{Ti}$) CFs at the instants of rupturing of the filaments and forms a potential barrier and persists until the gap widens beyond the tunneling probability.. The switching cycle was repeated over 1000 cycle to ensure the robust endurance feature of RRAM required for write, read, and erase operations as shown in Figure 2 and these tiny spikes of current consistently appears at the IRS states during the RESET process. The symmetric Pt electrode structure on both sides of the device is expected to ensure a uniform electric field distribution and minimize the electrode material interface.[35] Figure S4 shows the reset voltage as well as magnitude of HRS increased with thickness of SMTO layer. As the thickness increases, the electric field applied across the device becomes more distributed, leading to a lower local field at the interface, where switching occurs making it harder to initiate filament formation. This leads to higher resistance in thicker films resulting in larger power dissipation and joule heating, which impacts the stability of the switching process and requires higher voltages. The longer path for ion migration and filament growth within thicker films, further necessitates higher switching voltages. In summary, the IRS are believed to be the potential features of the bipolar RRAM devices for their utilization in neuromorphic computing, however, the spike observed are specifically in the IRS regions of the devices and potentially challenging the definition of a state in a given circuit. Hence, it is important to understand these tiny spikes of currents that repeatedly appear at the IRS regions of the reset process. To elucidate the formation of filaments and the nature of defects present in the devices, XPS studies were carried out on as-grown pristine samples and from the surfaces of the switched samples, from the regions in the vicinity of the electrode which also partially encompasses edges of the top electrodes.

### 3.3. Chemical Characterization

As detailed earlier, the XPS studies were carried out in both pristine and switched samples in the vicinity of the electrode encompassing the switched regions of the samples at different thicknesses and is shown in Figure 3. Figure 3 (a-h) shows the core level spectrum of Sr, Ti, Mn and Oxygen ions obtained from the as grown films of both pure STO and SMTO. The XPS spectrum revealed the presence of $Ti^{4+}$ ions which is one of the most stable states of Titanium in a $ABO_3$ perovskite structure as shown in Figure 3(b) and 3(f). [13,36] The oxygen core level spectrum shown in Figure 3(c)(STO) and 3(g) (SMTO25) clearly indicates the neutral oxygen peak at ~530 eV and the ~532 eV peak conventionally believed to arise from the doubly ionized oxygen vacancy ( $V_{\ddot{O}}$), which is recently reported as the spectrum arising from the adsorbed species of doubly ionized oxygen vacancy site.[37] These oxygen defects could be the naturally existing thermodynamic defects present in the as grown films, which are present in both STO and SMTO25. The peaks around 132.8 and 134.6 eV in Figure 3 (a and e) correspond to $3d_{5/2}$ and $3d_{3/2}$ of bulk $Sr^{2+}$ respectively. Besides these peaks, two minor peaks around 133.7 ± 0.1 and 135.5 ± 0.1 eV arise from the adsorbed species most likely hydroxyl ion that are adsorbed around the oxygen defects and interacts with $Sr^{2+}$ cations and resulting in the formation of $Sr(OH)_2$.[38,39,40] Figure 3(h) shows the very minor peak that arises from the Mn cation which is just 2 % of the whole system and hence a very weak signal could be detected from Mn cation core level spectrum. In spite of the weak signal, it could be evidently observed that the spectrum is indicating a predominant $Mn^{4+}$ state present in SMTO25 and the corresponding binding energies were tabulated in Table S1.

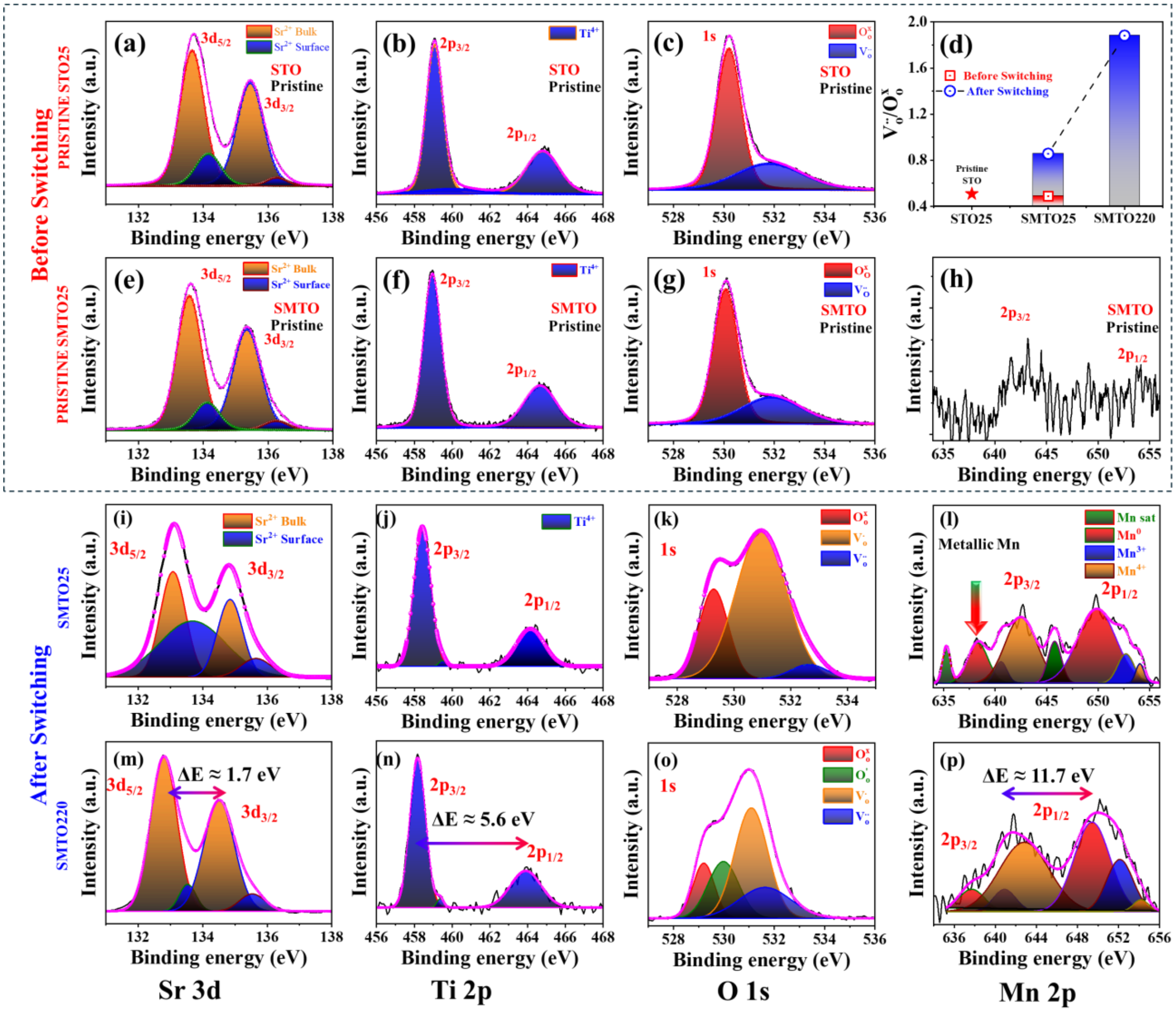

Figure 3. High resolution x-ray photoelectron core spectra of (a) Sr 3d, (b) Ti 2p, (c) O 1s from pristine STO25, (d) ratio of oxygen vacancy ($V_O^{\cdot\cdot}$) to lattice oxygen $O_O^x$ before and after switching

cycles, HR-XPS of (e) Sr 3d, (f) Ti 2p, (g) O 1s and (h) Mn 2p from pristine SMTO25 obtained before switching cycles and HR-XPS of (i and m) Sr 3d, (j and n) Ti 2p, (k and o) O 1s and (l and p) Mn 2p from (top) SMTO25 and (bottom) SMTO220 obtained after switching cycles.

Interestingly the XPS data collected in the vicinity of the switched electrodes which partially includes the electrode captures the details of the role of defect interplay in the switching characteristics of SMTO thin films. Figure 3 (i to p) shows the core level spectra of all the elements present, collected after electrical switching studies. The data clearly indicates the accumulation of different type of defects in the vicinity of the electrode. Very interestingly the Mn spectrum had shown a significant enhancement indicating various plausible Mn cations and metallic states in both SMTO25 and SMTO220 (~220nm thick). As we can evidently see that in addition to a weak metallic Mn signal, there is a dominant evolution of $Mn^{3+}$ signal probably giving rise to a singly negative Mn substitutional defect in the site of $Ti^{4+}$ ($Mn'_{Ti}$) and an associated enhancement of both singly ionized ($V_{\dot{O}}$) and doubly ionized ($V_{\ddot{O}}$) oxygen vacancy to maintain the charge neutrality. However, the titanium spectrum remains the same in both before and after switching studies. The oxygen 1s spectra after switching in Figure 3 (k and o) was split into four peaks arising from $O^{2-}$ (529.3 ± 0.1 eV), $O^{-}$ (530.8 ± 0.4 eV) and two vacant oxygen equivalent sites (531.32 ± 0.16 and 532.19 ± 0.37 eV)[12] due to hydroxyl bond formed with surface defects. The non-stoichiometry oxygen ($O^{-}$) sites in Figure 3 (o) (green) were introduced due to the chemical potential change in the STO matrix owing to applied electric field and Mn dopants at $Ti^{4+}$ sites. It is also worth noting that the rise of Mn signal indicates the accumulation of Mn based defects in the vicinity of top electrode and due to the mixed oxidation states of Mn there is an emergence of singly ionized oxygen vacancy. Since the charged oxygen vacancies at the surface is known to acquire some sorbed species the associated rise in Sr surface states could be evidently observed, which was very minor in the case of as grown samples. Most importantly, Figure 3 (d) shows the ratio of the area under the curves of doubly ionized oxygen vacancy to the neutral oxygen peak and it unambiguously demonstrates the accumulation of defects around the electrode and which is not predominant in the as grown samples. The existence of metallic manganese ($Mn^0$) in switched SMTO confirms the CFs formed by Mn ions accumulation that gets reduced by the electrons in the vicinity. Besides as the thickness of SMTO layer scales up, Mn spectra intensity drops suggests that $Mn'_{Ti}$ is accumulated at the surface of thinner films while in thicker films, it is buried underneath and the STO matrix play a significant role. Correlating the core spectra at different thickness from Figure 3, it is obvious that the effect of Mn doping is reduced with thickness increment and reduction in the relative amount of non-stoichiometry oxygen ($O^-$) introduced for local charge compensation. The thickness dependent CFs formation could be the resultant of variation in diffusivity of different possible defects migrated toward surface since the magnitude of electric field increases with reduced thickness for the corresponding voltages. XPS core spectra analysis of SMTO at different thickness offers insights into the prediction of conduction mechanism with surface Sr-O domination in lower (SMTO25) thickness could result in filamentary based switching characterized by abrupt set and reset. Whereas in intermediate (SMTO220) thickness as slight gradual set and reset process is expected owing to the reduction in surface Sr-O termination, presence of $Mn^{3+}$ in $Ti^{4+}$ sites and increased nonstoichiometric lattice oxygen ($O'_O$) sites. It also suggests that CFs could possibly be comprised of metallic Mn, $Ti_2O_3$ and SrO or a combination of these CFs with reduction in the influence of Mn doping at higher thickness.

From XPS studies, it is clear that Mn dopants play significant role in low (25 and 54nm) and mid (110 and 220nm) range thickness of SMTO. Furthermore, Figure 2 demonstrates multistate resistive switching behavior with the increment in thickness of SMTO characterized by the multiple distinct resistance jumps in both set and reset process. The possibility of presence of different types of CFs such as $V_O^{\cdot\cdot}$, $Mn^{3+}$ and $Sr^{2+}$ (Mn and SrO) based filaments is supported by Figure 2 and 3. The multiple resistive states at higher thickness in Figure 2 can be visualized as a parallel connection CFs[10] of different types of filaments formed at different set voltage based on the activation energy required for the formation of the corresponding CFs. The set voltage peaking at mid-range thickness in Figure S5 might be plausibly due to the formation TiO terminated surface. Furthermore, the decreased capacitance in a thicker layer can affect charge accumulation at the electrode filament interface, contributing to higher voltages needed to drive the resistive switching voltage with increasing thickness of SMTO. This uniformity in electrical behavior ensures repeatable switching, and maintains stable LRS and HRS states, which are crucial for reliable RRAM applications.

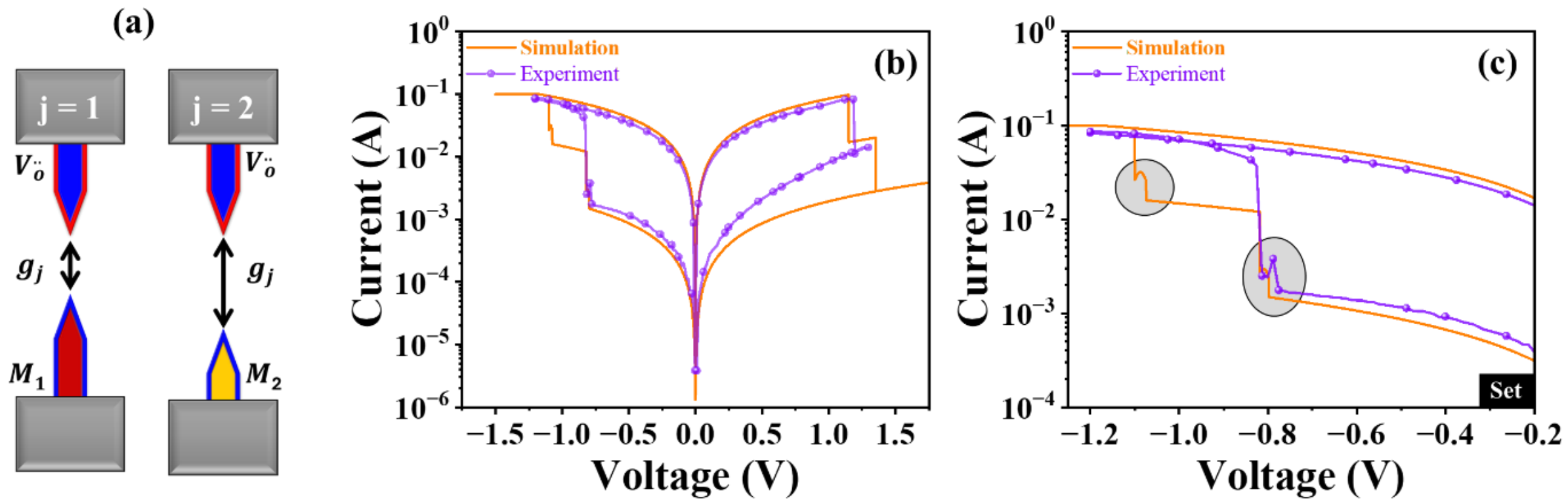

Figure 4. (a) Graphical representation of two parallel $p^+$-$n^+$ like junctions formed in SMTO between Pt electrodes with unique gap kinetics of each individually, (b) Resistive switching curve of SMTO25 obtained by simulation (orange) and experiment (purple) and (c) zoomed view of semi-log I-V characteristics of Pt/SMTO25/Pt device highlighting the intermediate tunnelling process during set.

To unveil the linkage among brief distribution in set voltage for consecutive cycles and dynamic tunneling between heavily doped CFs of RRAM, numerical simulation was rendered by integration of leakage current via SMTO matrix and filament current flowing through CFs. As pointed out above, these CFs act as $p^+/n^+$ doped donors and acceptors and propagate towards opposite polarity narrowing down the gap between the electrodes analogous to Esaki junction with a physical gap of SMTO in place of depletion layer with no negative differential resistance.[41] A compact device model is proposed for the pre-contact tunnel current spike fitted to the peak voltage and shape exponent function.

The model was designed with two parallel junctions indexed by $j$ between CFs and forms simultaneously as visualized in Figure 4(a). Each junction carries the gap ($g_j$) between the two approaching tips, closing at the combined rate of both opposing electrodes, with contact at $g_{min}$ = 0.4 nm (close to one unit cell of STO). For the degenerate junction of each CFs, the tunneling spike peak voltage ($V_{p,j}$) was derived from the degeneracy energies ($\xi$(N)) of both sides[41] as described by Equation 1 and 2

$$\xi(\mathrm{N}) = k_B T \left[ ln\left(\frac{N}{N_{ref}}\right) + \frac{1}{\sqrt{8}}\frac{N}{N_{ref}} \right] \tag{1}$$

$$V_{p,j} = \frac{\xi(\mathrm{N}_d)+\xi(\mathrm{N}_{a,j})}{3}, V_{esaki,j} = \frac{V_{p,j}}{\alpha} \tag{2}$$

Here, $\alpha$ is a fitting parameter. Since $V_o^{\cdot\cdot}$ is the same opposing filament for both metal ion junctions, the donor concentration ($\mathrm{N}_d$) was expressed by a single term for both junctions. However, the presence of $Ti_2O_3$ and metallic Mn as noticed from XPS analysis, suggests the formation of multiple elementary metal ion CFs. Hence, the acceptor concentration ($\mathrm{N}_{a,j}$) in each CFs was distinguished for each individual junction (j) likely $Ti^{3+}$ and/or Mn ions. It is to be noted that these carrier concentrations represent the concentration in CFs and not the SMTO matrix.

Despite fixing the gap between CFs, the gap dynamic was expressed by a piecewise drift law in Equation 3[42] with dead zones (LRS) between $V_{set,j}$ and $V_{reset,j}$.The dead zone supplies the memory between threshold voltages and the gap is frozen.

$$\frac{dg_j}{dt} = \begin{cases} -k_j \times \left(\frac{V}{V_{d,j}} - 1\right) & , if\ V \leq -V_{d,j} \\ +k_j \times \left(\frac{V}{V_{reset}} - 1\right) & , \quad V \geq V_{reset,j} \\ 0 & , \quad otherwise \end{cases} \tag{3}$$

Since the tip of the CFs are expected to advance towards each other, the diffusion coefficient ($k_j$) is the combined closure rate rather than a single filament velocity and $V_{d,j}$ represents the onset of the migration of the defect ions. A fixed step was utilized to perform the integration with clipping the drift range ($g_{max}$) as in Equation 4.

$$g_j^{(n+1)} = clip\left[g_j^{(n)} + \frac{dg_j}{dt}\Delta t\ ;\ g_{min}, g_{max}\right] \tag{4}$$

$g_j^{(n+1)}$ is not allowed to take values equal to the as-deposited film thickness, instead, the data curation were steady-state cycles, in which the gap only moves between $g_{min}$ and $g_{max}$. Later the defect migration kinetics takes over the RESET behavior of the device. The gap-dynamics model used here is based on the unified current-voltage model by Harry Chung et al[42], which uses fitting parameters ($k_{on}$, $k_{off}$) to scale the rate of change of the filament boundary for a unidirectional growth.[42] Since the devices studied here is expected to have filaments forming from both electrodes simultaneously, we define separate rate parameters $k_j$ for each filament. Here, $k_j$ reflects the combined contribution of both filament boundaries to gap closure, rather than that of a single filament towards a stationary electrode. This simulation should therefore be interpreted as a numerical proof of concept rather than a physically accurate depiction of bidirectional filament dynamics, which requires molecular dynamics-based simulation, which is beyond the scope of this study.

The total current measured was proposed to be the summation of the current ($I_L(V)$) offered by Poole-Frenkel (first term of Equation 5) and filament current ($I_j(V)$) (second term of equation 5) dictated by the summation of tunneling and ohmic conduction aided by the filaments.[43] The form of the electrical transport of the individual mechanisms are expressed in Equation 6, 7 and 8 with the characteristic switching voltage ($V_{p,j}$)[41] for each filament species was derived from Esaki junction model of the heavily doped filament junctions.

$$I(V) = I_L(V) + \sum_j I_j(V)\,, |I| \leq I_{comp} \tag{5}$$

$$I_L(V) = A_L V e^{(B_L\sqrt{|V|})} \tag{6}$$

$$I_j = \begin{cases} \frac{V}{R_{s,j}} & , \quad if\ g_j \leq g_{min} \\ A_j x_j e^{m(1-x_j)} \mathcal{T}(g_i) & , \quad g_{min} \leq g_j \leq g_{tun} \\ 0 & , \quad g_j > g_{tun} \end{cases} \tag{7}$$

$$x_j = \frac{\alpha.|V|}{V_{p,j}}\,, \mathcal{T}(g_i) = e^{-2\kappa(g-g_{\min})}, \kappa = \frac{\sqrt{2m^*\phi}}{\hbar} \tag{8}$$

No current is assumed to flow via CFs, unless $g_i$ is less than $g_{tun}$. Equation 7(b) is the Esaki form raised to a compression exponent 'm', multiplied by the Wentzel–Kramers–Brillouin (WKB) transmission of the residual gap. The exponent does not move the peak, which is fixed at $x_j$= 1 for any m (m=1 for band-to-band tunneling), but it narrows the negative-resistance region.

Once $g_i$ reaches $g_{min}$, ohmic conduction dominates the tunnel current on top of the Poole-Frenkel conduction inherently present in the system. Equation (7) represents the conduction by means of completed junction beyond its valley, limited by the series resistance of the filaments. The branch was restricted to V < 0 and to junctions that have not yet formed.

The RS curve for SMTO was simulated for an oxide layer thickness of 25 nm and $g_{tun}$ of 0.6 nm as was overlayed with I-V curve of SMTO25 generated from experiment as shown in Figure 4(b). An iterative fitting was carried out between experimental and simulation data to derive the device conduction parameters and were tabulated in Table S2. Resistance in both HRS and LRS agree with the experiment data to better than 0.4%. Examining precisely at the switching voltage as displayed in Figure 4(c), both simulation and experiment showcases a current spike marked by grey circle shortly preceding $V_{set}$. This spike in current is expected to arise due to the tunneling of electron form $n^+$-doped CFs to $p^+$-doped CFs prior to complete filament formation. It is to be noted explicitly that switching from HRS to LRS in RRAM is not due to tunneling rather because of ohmic conduction by the completely formed CFs. Hence, tunneling is an intermediate conduction mechanism dominated in $g_{min} \leq g_j \leq g_{tun}$ regime which potentially occurs at the IR states of the device.

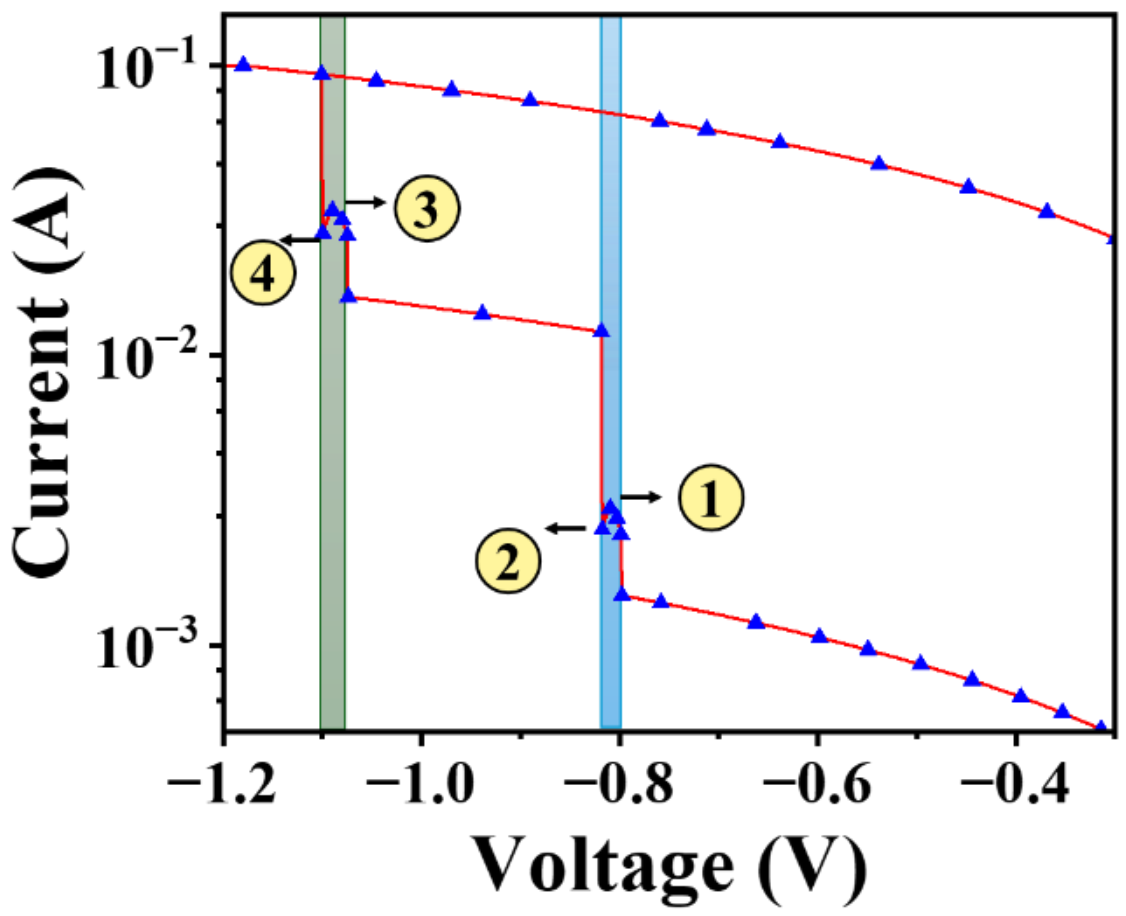

Figure 5. Tunneling dominated voltage windows for the simulation SET process of SMTO25. Labels 1 and 3 are the tunnel peaks and 2 and 4 are the CFs formation steps.

The deconvolution of various conduction mechanisms in different voltage regimes separated by integration of the gap dynamics of Equation 3. By integration of gap dynamics of HRS in Equation 3 within $g_{min}$ and $g_{max}$, the threshold voltage range between initiation of electromigration of defect and complete filament formation can be segregated as in Equation 9

$$\left(|V_{set,j}| - V_{d,j}\right)^2 = \frac{2V_{d,j}(g_{max}-g_{min})}{k_j\beta} \tag{9}$$

where β is the voltage sweep rate. Further, onset voltage ($V_{open_j}$) of the tunneling and it's dominated conduction voltage window ($w_j$) was derived by integration of Equation 3 in the limits of $g_{min}$ and $g_{tun}$ as in Equation 10, 11 and displayed in Figure 5,

$$|V_{open\,j}| = |V_{set\,j}| - w_j \tag{10}$$

$$w_j = \frac{(g_{tun} - g_{min})\left(|V_{set\,j}| - V_{d\,j}\right)}{2(g_{max} - g_{min})} \tag{11}$$

The peak bias voltages ($V_{p\,j}$)were fixed by the filament carrier densities, once α was chosen. For instance, the carrier concentration calculated by iterative fitting of simulated curve and experiment data are in the order of $10^{21}$ ~$10^{22}$ cm$^{-3}$. This high carrier concentration supports the claim in the above section that though pristine SMTO is lightly doped (2%), CFs are localized and expected to behave similar to heavily doped semiconductor, in some cases metallic. These defects were homogeneously distributed in pristine SMTO that has migrated and accumulated to form the CFs.

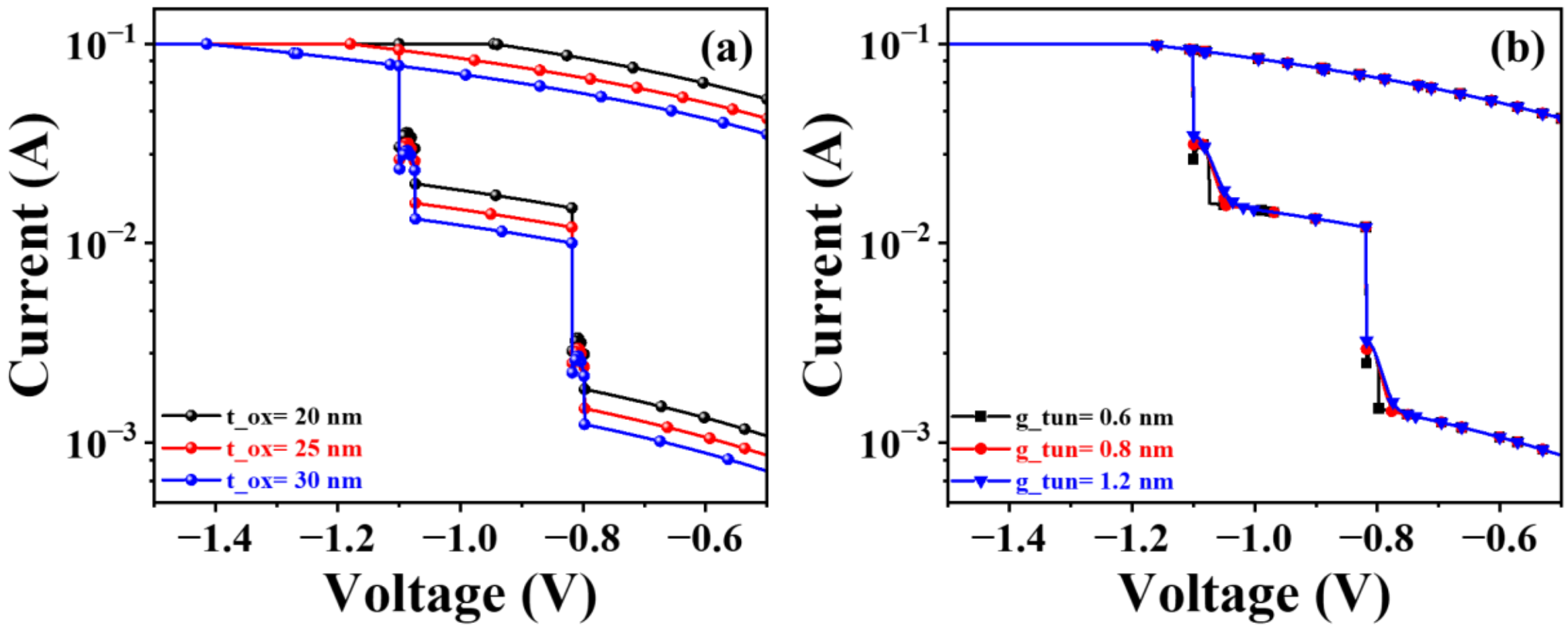

Figure 6. Semilog I-V curve of simulation of set process with (a) 0.6 nm tunneling onset gap but differing oxide thickness and (b) fixed oxide thickness of 25 nm with changing tunneling onset gap.

From the learnings of the above observations, the film thickness and the tunneling onset gap were varied in simulation and plotted in Figure 6. Upon increasing the thickness in Figure 6(a), the magnitude of resistance in both HRS and LRS were found to increase. Interestingly, this observation coincides with experimental results as noted in Figure 2 without significant variation in ON/OFF ratio. Besides, the set voltage seems to be independent of the total thickness of the film suggesting that it relies on the nature of the CFs. Variation in onset tunneling gap in Figure 6(b) shows no influence in magnitude of HRS or LRS. However, the magnitude of tunneling current spike is proportional to the onset tunneling gap and decides the set voltage. It could be hypothesized that the variation in set voltage from preceding to succeeding cycle is due to the local fluctuation in tunneling barrier energy in different orientation of the polycrystalline columnar grains, grain boundaries. These fluctuations lead to variations in tunneling onset gap resulting in distribution in $V_{set}$. In addition to electromigration of defects, reset process is also dominated by Joule's heating in certain systems, which is not focused and is the limitation in this model.

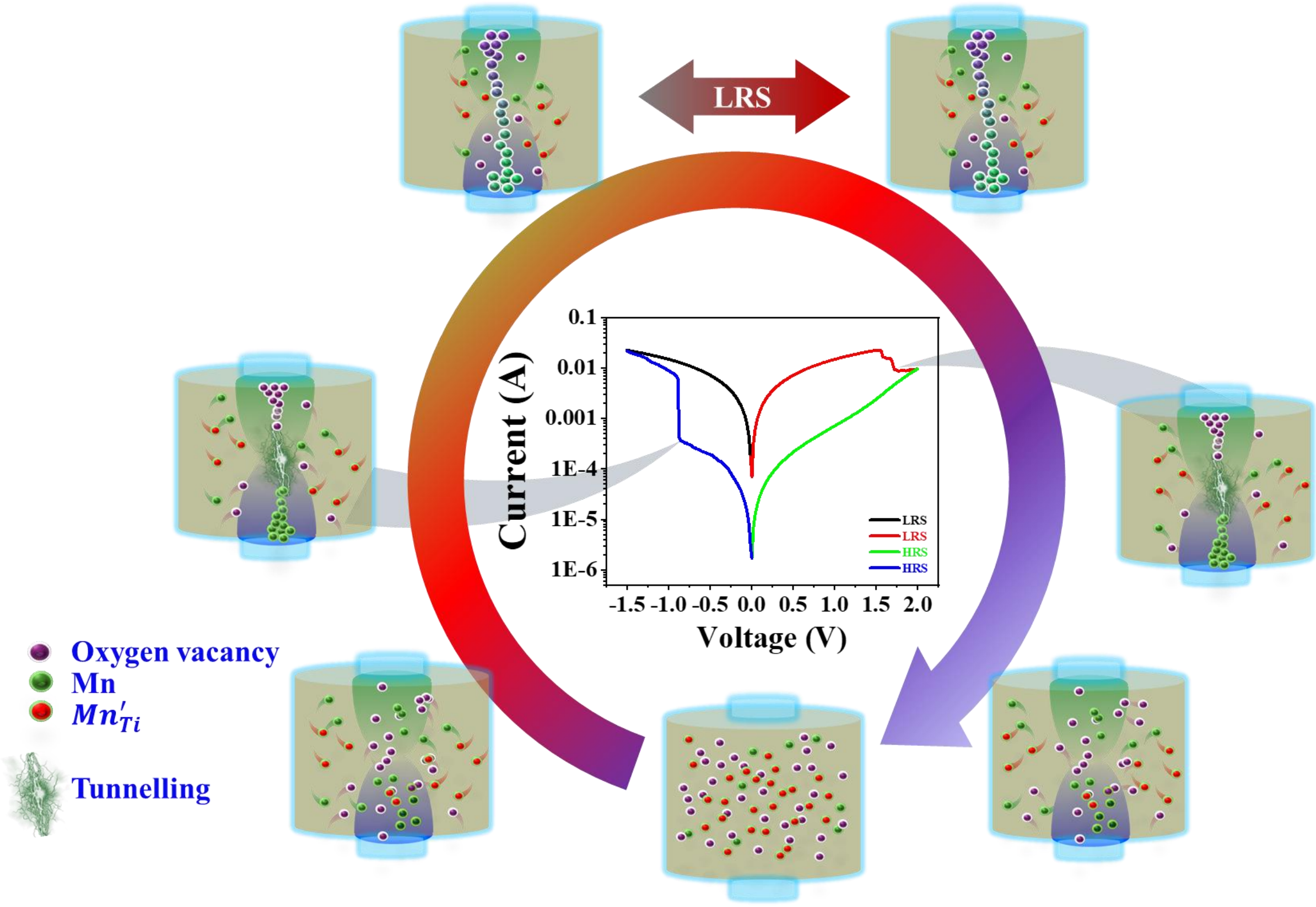

Figure 7. Thematic Figure of resistive switching mechanism of SMTO870

The transition between LRS and HRS in RRAM and corresponding plausible defect dynamics is shown in as shown in the thematic Figure 7. To summarize, charged defects generated by $Mn^{3+}$ doping at $Ti^{4+}$ sites ($Mn'_{Ti}$) and oxygen vacancies were homogeneously distributed before switching. Under applied electric field, electromigration of Mn ions and oxygen vacancy gets initiated and accumulated at the Pt electrode from its vicinity. During electromigration under negative bias, $V_O^{\cdot\cdot}$ and $V_O^{\cdot}$ move towards top electrode. A tiny fraction either $V_O^{\cdot\cdot}$ and/or $V_O^{\cdot}$ could have interacted with $Mn'_{Ti}$ in the proximity to form metallic Mn. At the instant prior to switching, tunneling occurs between oppositely charged CFs and followed by the conduction via CFs with low resistance obeying Ohms law. Increase in voltage in positive bias beyond reset voltage dissociates CFs and restore HRS as Poole- Frenkel conduction takes over. Its worth mentioning that multiple IRS states at discrete voltage windows were also observed amidst the reset process.

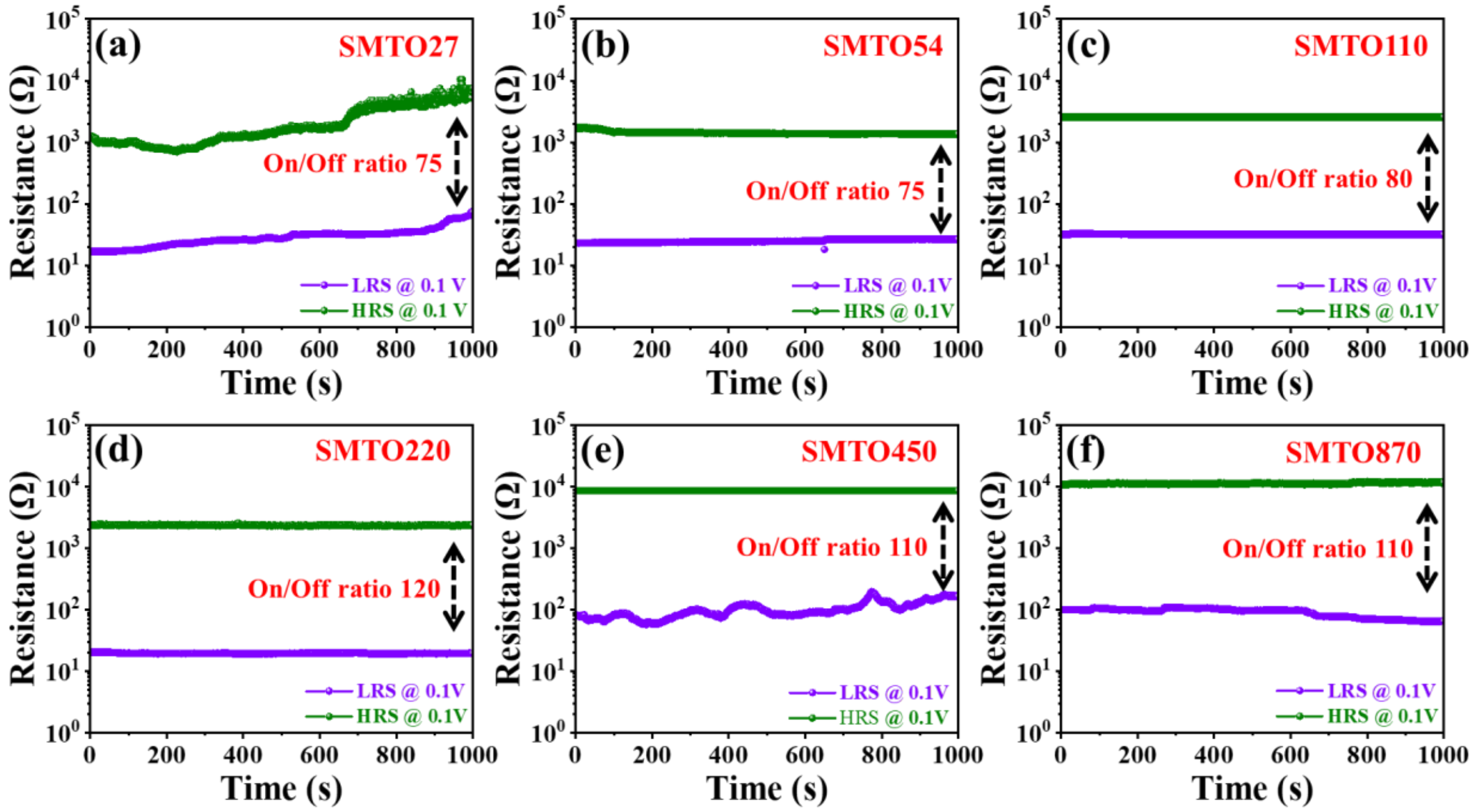

Figure 8. Retention characteristics of Pt/SMTO/Pt device with different SMTO thickness of (a) 25nm, (b) 54nm, (c) 110nm, (d) 220nm (e) 450nm and (f) 870nm

Inspite of the conduction mechanisms that involve in determination of the IRS of SMTO films the overall ON/OFF ratio of the devices were analyzed via retention studies. Retention studies were conducted to assess the long-term stability of the devices to maintain non-volatility in stored data. Retention is evaluated by monitoring the resistance ratio between HRS and LRS and was conducted over $10^3$ seconds at a read voltage of 0.1 V and is shown in Figure 8. It is worth noting that all the devices of SMTO studied show excellent retention features hence proving to be a potential candidate for RRAM based devices.

4. Conclusion

In summary, defect engineered Mn doped STO films show promising features of resistive switching. The conductive filament comprises of negatively charged $Mn'_{Ti}$ defects and positively charged $V_O^{\cdot\cdot}$ and $V_O^{\cdot}$ defects resulting into bipolar switching with forming free reduced set voltage(s). XPS core spectrum analysis obtained from before and after switching of the device confirms the lateral migration of the defects towards the biased electrodes and indicates the plausible defects beneath the electrodes that would form the CFs. The dynamic tunnel process is expected to happen as a intermediate conduction at the instant of connecting and snapping of the filament during the set and reset operations respectively. The numerically simulated modified conduction equations with inclusion of tunnel barrier clearly demonstrates the possibility of such a conduction process between the ON/OFF states of the devices. More precisely the dynamic tunneling is predominantly observed in the IR states of the thicker films and prior to the transition states of the thinner films. The numerically simulated curve with inputs from experimental observations show remarkable coincidence with the experimental observed switching curves indicating the presence of intermediate dynamic tunneling mechanism at prior to formation and at the instant of detachment of CFs. In addition, all fabricated devices showcase promising forming-free resistive switching with ON/OFF ratio of $10^2$ measured across $10^3$ cycles exhibiting a non-volatile memory retention beyond $10^3$s.

ASSOCIATED CONTENT

**Supporting Information** are available free of charge.

Supporting Information includes the target synthesis procedure with grazing incidence x-ray diffraction of target and thin films of SMTO are available. The binding energy (eV) obtained from XPS spectra, empirical constants used, iterative fitting parameters calculated from two-junction filaments model are tabulated. Semilog I-V curve of SMTO thickness of (a) 54nm, (b) 110nm, (c) 870nm are also available.

AUTHOR INFORMATION

**Corresponding Author**

*Ranjith Ramadurai (E-mail: ranjith@msme.iith.ac.in).

**Present Addresses**

†Institute of Physical Chemistry, Polish Academy of Sciences, Kasprzaka 44/52, Warsaw, 01-224 Poland.

Author Contributions

‡Gowtham Velpandi and Ishaan Tushir contributed equally. Gowtham Velpandi and Durairaj Murugan fabricated, performed structural and electrical characterization of the RRAM devices. Ishaan Tushir was involved in numerical simulation of the two-junction filaments model. Gowtham Velpandi and Thanumalayaperumal Thirunavukarasu involved in chemical characterization. Gowtham Velpandi and Ishaan Tushir wrote the first version of the manuscript. Ranjith Ramadurai conceptualized, designed, acquired funds, reviewed and wrote the final version of the manuscript. The manuscript was written through contributions of all authors. All authors have given approval to the final version of the manuscript.

**Funding Sources**

The authors would like to thank DST-FIST Program (Grant no. SR/FST/ETI-421/2016) for Microstructural Characterization and Department of Science and Technology (Grant No: DST/ SERB/EMR/2017/003159/MMM) for financial assistance to carryout studies.

## ACKNOWLEDGMENT

Author GV would like to thank Ministry of Education (MoE), India for the fellowship and author DM would like to acknowledge Indian Institute of Technology Hyderabad (IITH) for Institute Postdoctoral Fellowship (IPDF).

## REFERENCES

---

[8] Fakhry, D.; Abdelsalam, M.; El-Kharashi, M. W.; Safar, M. A Review on Computational Storage Devices and near Memory Computing for High Performance Applications. *Memories - Materials, Devices, Circuits and Systems* **2023**, *4*, 100051. https://doi.org/10.1016/j.memori.2023.100051.

[9] Fedotov, M. I.; Klyukina, E. V.; Vankaev, A. S.; Znamenskii, A. G.; Koveshnikov, S. V. Experimental and Theoretical Study of Multifilamentary Resistive Switching in Nanoscale Transition Metal Oxide Films. *ACS Omega* **2025**, *10* (37), 43043–43052. https://doi.org/10.1021/acsomega.5c06052.

[10] Asapu, S.; Maiti, T. Multifilamentary Conduction Modeling in Transition Metal Oxide-Based RRAM. *IEEE Trans. Electron Devices* **2017**, *64* (8), 3145–3150. https://doi.org/10.1109/TED.2017.2709249.

[11] Koehl, A.; Wasmund, H.; Herpers, A.; Guttmann, P.; Werner, S.; Henzler, K.; Du, H.; Mayer, J.; Waser, R.; Dittmann, R. Evidence for Multifilamentary Valence Changes in Resistive Switching SrTiO3 Devices Detected by Transmission X-Ray Microscopy. *APL Materials* **2013**, *1* (4), 042102. https://doi.org/10.1063/1.4822438

[12] Murugan, D.; Velpandi, G.; Ramadurai, R. Robust Filament Formation via Defect Engineering in Mn: SrTiO3 for Oxygen Vacancy Driven RRAM Device. *Materials Letters* **2026**, *404*, 139571. https://doi.org/10.1016/j.matlet.2025.139571.

[13] Yang, Y.; Cao, Z.; Mao, S.; Qin, J.; Rao, Z.; Liu, M.; Ke, C.; Zhao, Y.; Sun, B. Effect of Heterojunction Order between CaTiO3 and Mn Doped SrTiO3 on Memristive Performance and Its Mechanism Analysis. *Applied Materials Today* **2023**, *35*, 101938. https://doi.org/10.1016/j.apmt.2023.101938.

[14] Park, G.-S.; Li, X.-S.; Kim, D.-C.; Jung, R.-J.; Lee, M.-J.; Seo, S. Observation of Electric-Field Induced Ni Filament Channels in Polycrystalline NiOx Film. *Applied Physics Letters* **2007**, *91* (22), 222103. https://doi.org/10.1063/1.2813617.

[15] Tang, M.; Dai, L.; Cheng, M.; Zhang, Y.; Wang, Y.; Zhong, X.; Wang, J.; An, F.; Ma, M.; Huang, M.; Li, C.; Li, J.; Zhong, G. High-Throughput Screening Thickness-Dependent Resistive Switching in $SrTiO_3$ Thin Films for Robust Electronic Synapse. *Adv Funct Materials* **2023**, *33* (23), 2213874. https://doi.org/10.1002/adfm.202213874.

[16] Song, L. W.; Niu, S. S.; Sun, Y. C.; Hua, L. F.; Zhao, X.; Chen, W. Crystallinity Dependence of Resistive Switching in Ti/$Pr(Sr_{0.1}Ca_{0.9})_2Mn_2O_7$/Pt: Filamentary versus Interfacial Mechanisms. *Appl. Phys. Lett.* **2014**, *104* (9), 093502. https://doi.org/10.1063/1.4867483.

[17] Cooper, D.; Baeumer, C.; Bernier, N.; Marchewka, A.; La Torre, C.; Dunin-Borkowski, R. E.; Menzel, S.; Waser, R.; Dittmann, R. Anomalous Resistance Hysteresis in Oxide ReRAM: Oxygen Evolution and Reincorporation Revealed by In Situ TEM. *Advanced Materials* **2017**, *29* (23), 1700212. https://doi.org/10.1002/adma.201700212.

Supporting Information

# Dynamic Tunneling a Prior Phenomenon to Intermediate Resistive States in Defect Engineered Mn-$SrTiO_3$ RRAM Devices

*Gowtham Velpandi‡, Ishaan Tushir‡, Durairaj Murugan†, Thanumalayaperumal Thirunavukarasu and Ranjith Ramadurai**

Department of Materials Science and Metallurgical Engineering, Indian Institute of Technology Hyderabad, Kandi, Sangareddy, Telangana 502285, India.

Target preparation

2% Mn-doped $SrTiO_3$ (SMTO) powder was synthesized through a solid-state reaction method, involving a multi-step process. Initially, precursor powders of $SrCO_3$, $TiO_2$ and $MnO_2$ were meticulously weighed and mixed in a predetermined molar ratio, ensuring a homogeneous dopant distribution. The mixture was then subjected to ball milling at 150 rpm for 12 hours, employing a planetary ball mill, to achieve a uniform particle size and distribution. The mechanical alloying process facilitated the intimate mixing of the precursors, promoting a reaction between the constituent phases. Subsequently, the ball milled mixture was calcined at 1150 $^{0}$C for 4 hours in a high temperature furnace, under air atmosphere, to decompose the precursors and form the desired perovskite phase. Following calcination, the resulting powder was compacted into pellets at 40 kg/cm$^3$ using a hydraulic press, ensuring a high green density. Finally, the pellets were sintered at 1450 $^{0}$C for 5 hours in a furnace, promoting densification and crystallization of the SMTO ceramics.

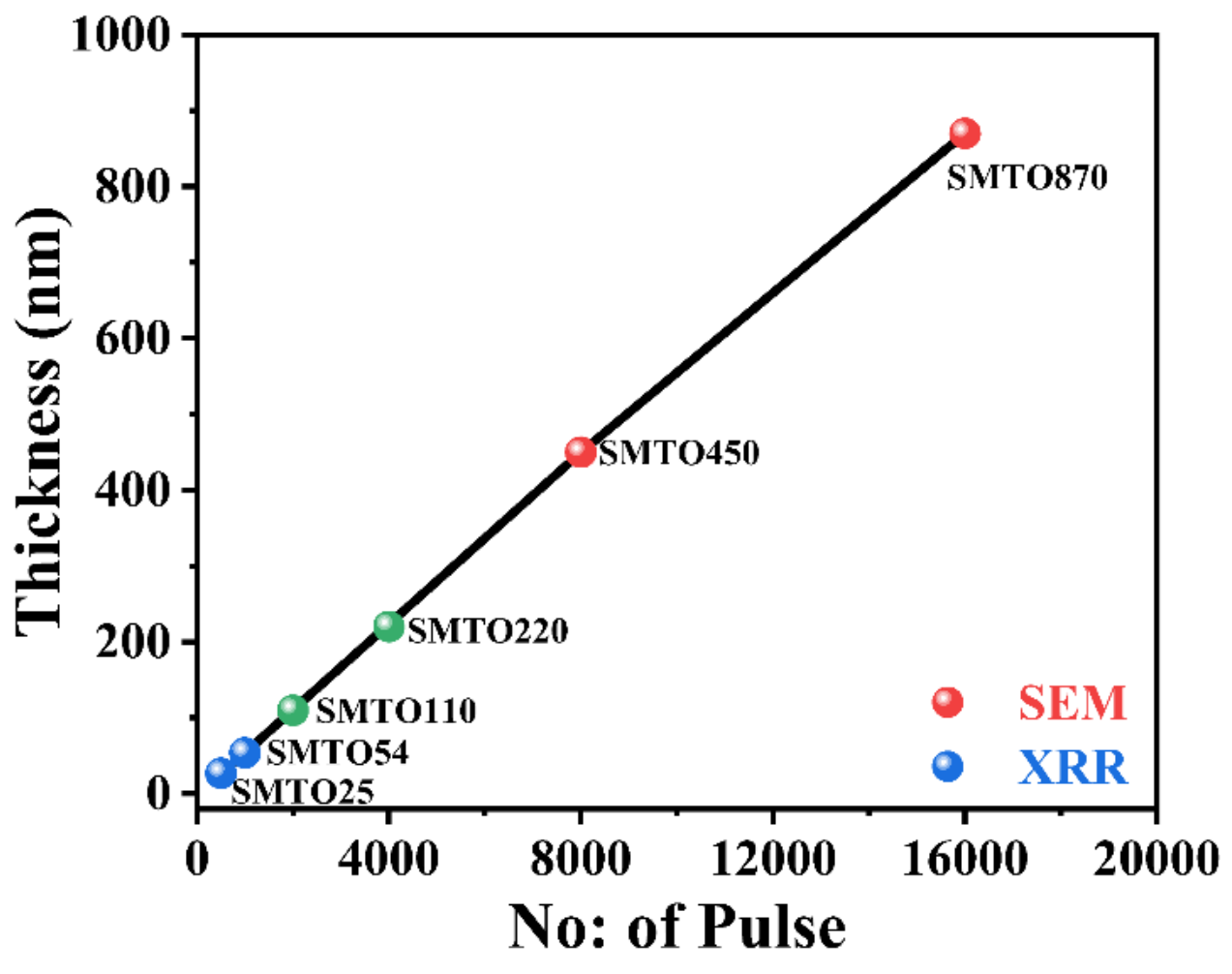

**Figure S1.** Growth rate optimization of SMTO

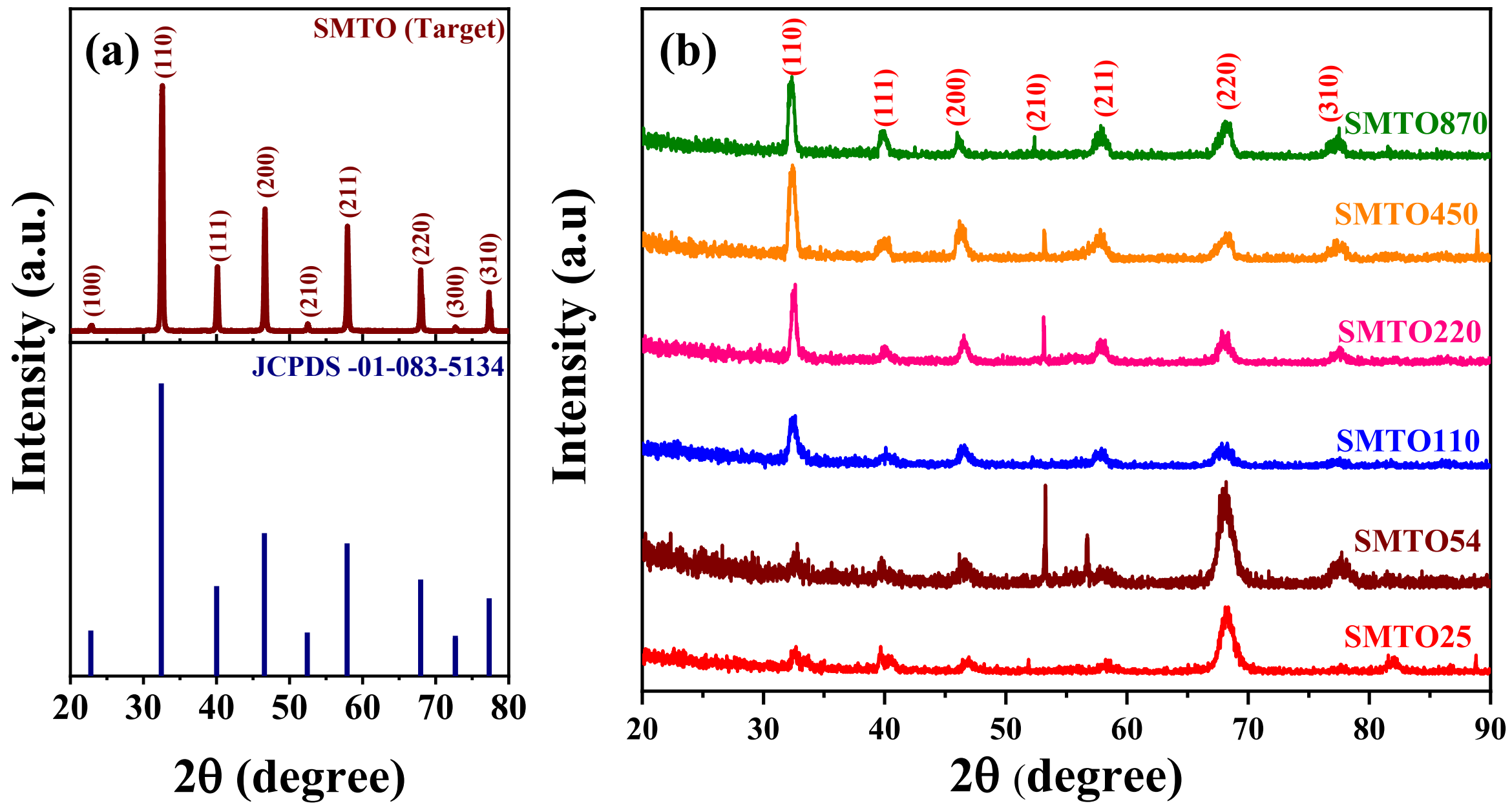

**Figure S2**. Grazing incidence x-ray diffraction of SMTO (a) target (up) and (b) thin films with varying thickness grown on a Pt-Si substrate.

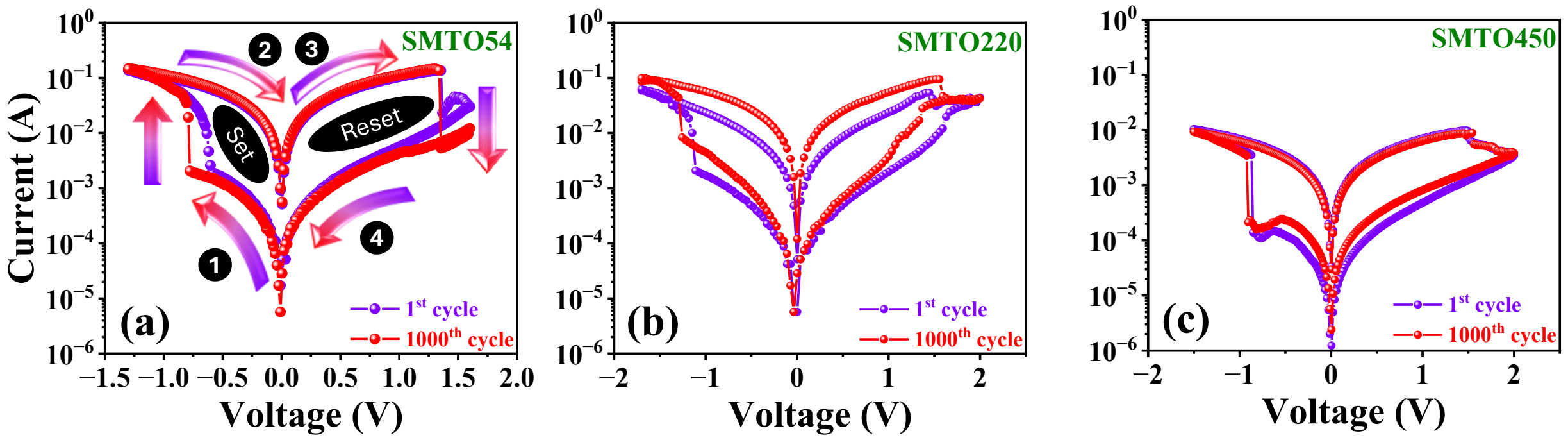

**Figure S3.** Semilog I-V characteristics of Pt/SMTO/Pt device with different SMTO thickness of (a) 54nm, (b) 110nm and (c) 870nm

**Table S1.** Binding energy (eV) obtained from deconvolution of the XPS spectra of SMTO25 and SMTO220 after switching.

| Sr 3d | Spin-orbit Splitting | $3d_{5/2}$ | | $3d_{3/2}$ | | ΔE | |
|---|---|---|---|---|---|---|---|
| | Oxidation State | $Sr^{2+}$ (Bulk) | $Sr^{2+}$ (Surface) | $Sr^{2+}$ (Bulk) | $Sr^{2+}$ (Surface) | $Sr^{2+}$ (Bulk) | $Sr^{2+}$ (Surface) |

| | | | | | | |
|---|---|---|---|---|---|---|
| SMTO25 | 133.08 ± 0.00 | 133.68 ± 0.13 | 134.85 ± 0.02 | 135.65 ± 0.16 | 1.77 | 1.97 |
| SMTO220 | 132.80 ± 0.01 | 133.53 ± 0.02 | 134.51 ± 0.01 | 135.53 ± 0.09 | 1.71 | 2.00 |
| Average | 132.83 ± 0.17 | 133.68 ± 0.10 | 134.62 ± 0.15 | 135.54 ± 0.07 | 1.79 | 1.86 |

| Ti 2p Spin-orbit Splitting | $2p_{3/2}$ | | $2p_{1/2}$ | | ΔE | |
|---|---|---|---|---|---|---|
| Oxidation State | $Ti^{3+}$ | $Ti^{4+}$ | $Ti^{3+}$ | $Ti^{4+}$ | $Ti^{3+}$ | $Ti^{4+}$ |
| SMTO25 | 458.42 ± 0.01 | 459.51 ± 0.08 | 464.18 ± 0.03 | | 5.75 | |
| SMTO220 | 458.18 ± 0.01 | 459.38 ± 0.06 | 463.90 ± 0.03 | | 5.80 | |
| Average | 458.19 ± 0.15 | 459.30 ± 0.19 | 463.78 ± 0.40 | 463.21 ± 0.93 | 5.58 | 3.91 |

| Mn 2p Spin-orbit Splitting | $2p_{3/2}$ | | | $2p_{1/2}$ | | | (ΔE) | | |
|---|---|---|---|---|---|---|---|---|---|
| Oxidation State | $Mn^{0}$ | $Mn^{3+}$ | $Mn^{4+}$ | $Mn^{0}$ | $Mn^{3+}$ | $Mn^{4+}$ | $Mn^{0}$ | $Mn^{3+}$ | $Mn^{4+}$ |
| SMTO25 | 638.3 ± 0.1 | 640.6 ± 0.1 | 642.5 ± 0.2 | 649.9 ± 0.1 | 652.7 ± 0.1 | 654.1 ± 0.1 | 11.6 | 12.1 | 11.6 |
| SMTO220 | 637.7 ± 0.1 | 640.9 ± 0.1 | 642.8 ± 0.1 | 649.4 ± 0.0 | 652.1 ± 0.1 | 654.2 ± 0.2 | 11.7 | 11.2 | 11.4 |
| Average | 638.0 ± 0.3 | 640.7 ± 0.2 | 642.7 ± 0.2 | 649.6 ± 0.2 | 652.4 ± 0.3 | 654.1 ± 0.1 | 11.6 | 11.7 | 11.5 |

| O1s Oxidation State | $O_o^x$ | $O_o^{\cdot}$ | $V_o^{\cdot}$ | $V_o^{\cdot\cdot}$ |
|---|---|---|---|---|
| SMTO25 | 529.35 ± 0.004 | - | 531.31 ± 0.010 | 532.24 ± 0.025 |
| SMTO220 | 529.20 ± 0.074 | 529.98 ± 0.275 | 531.09 ± 0.151 | 531.63 ± 1.869 |
| Average | 529.21 ± 0.09 | 529.98 ± 0.275 | 531.32 ± 0.16 | 532.19 ± 0.37 |

**Table S2. Empirical constants and iterative fitting parameters used in the two-junction filaments model.**

| Symbol | Meaning | $J_1$ | $J_2$ |
|---|---|---|---|
| $g_{max}$ | Drift range (post-reset gap) | 3.5 nm | 3.5 nm |
| $g_{min}$ | Contact gap | 0.4 nm | 0.4 nm |
| $g_{tun}$ | Tunnelling onset gap | 0.80 nm | 0.80 nm |
| $\alpha$ | Fitting parameter | 0.5 | 0.5 |
| $\beta$ | Voltage sweep rate | 2 | 2 |
| m | NDR compression factor | $1.47\times10^{4}$ | $1.50\times10^{4}$ |
| $\phi$, κ | Barrier height | 0.20 eV, 1.619 $nm^{-1}$ | ____ |
| $N_d$ | Carrier concentration in oxygen vacancy CF | $4.92\times10^{21}$ $cm^{-3}$ | $4.92\times10^{21}$ $cm^{-3}$ |
| $N_a$ | Carrier concentration in Metal ion CF | $6.03\times10^{21}$ $cm^{-3}$ | $1.04\times10^{22}$ $cm^{-3}$ |
| $\xi_d$ | Donor degeneracy energy | 0.550 eV | 0.550 eV |
| $\xi_a$ | Acceptor degeneracy energy | 0.657 eV | 0.657 eV |
| $R_s$ | Filament resistance | 77.9 Ω | 14.30 Ω |
| $m^*$ | Effective mass | 1.8 $m_e$ | 1.6 $m_e$ |
| $k$ | Combined gap closure rate | 1.618 | 1.449 |
| $A_L, B_L$ | Leakage current fitting parameter | $1.3\times10^{-3}$ , 0.4 | ____ |
| $A_j$ | Tunneling pre-factor fitting parameter | $2.14\times10^{-3}$ | $2.31\times10^{-2}$ |
| $I_{comp}$ | Compliance current | 0.1 A | 0.1 A |
| $w_j$ | Tunneling voltage window | 39.9 mV | 51.7 mV |
| $V_{peak}$ | placed at measured bias | −0.805 V | −1.083 V |
| $V_p$ | Junction bias, $\alpha V_{peak}$ | 0.4025 V | 0.5415 V |
| $V_d$ | Ion migration onset voltage | -0.20 V | -0.30 V |
| **Symbol** | **Meaning** | **$J_1$** | **$J_2$** |
| $V_{set}$ | Set voltage | −0.819 V | −1.101 V |
| $V_{reset}$ | Reset Set voltage | 1.35 V | 1.15 V |

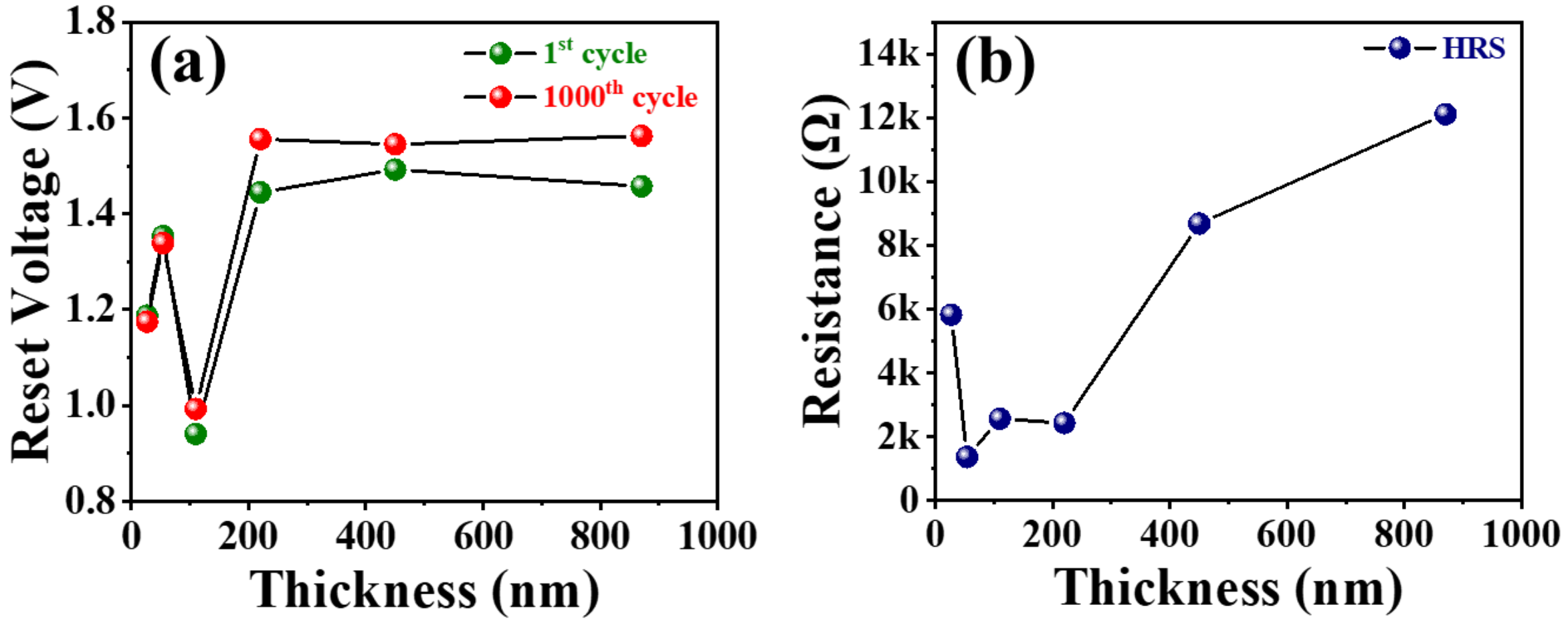

**Figure S4.** Variation in (a) reset voltage and (b) resistance in HRS at 0.1V with thickness.

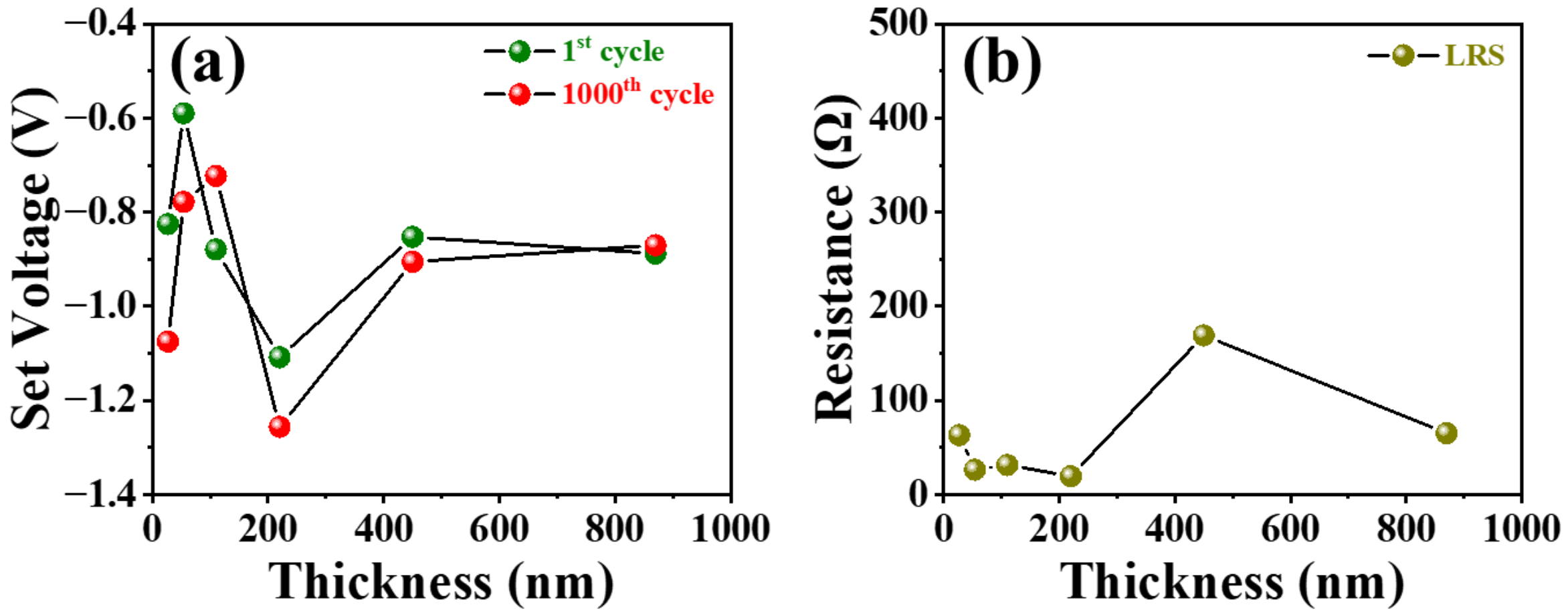

**Figure S5.** Variation in (a) set voltage and (b) resistance in LRS at 0.1V with thickness.